\documentclass[10pt,aps,prx,twocolumn,amsmath,amssymb]{revtex4-2}
\usepackage{mdframed}
\usepackage{graphicx}
\usepackage{dcolumn}
\usepackage{bm}
\usepackage{fancybox}
\usepackage{amsmath}
\usepackage{enumerate}
\usepackage{pslatex}
\usepackage{relsize}
\usepackage{setspace}
\usepackage{placeins}
\usepackage[utf8]{inputenc}

\usepackage[T1]{fontenc}
\usepackage{lmodern}
\usepackage{hyperref}
\usepackage{verbatim}
\usepackage{amssymb}
\usepackage[toc,page]{appendix}
\usepackage[version=3]{mhchem}
\usepackage{chemfig}
\usepackage{numprint}
\usepackage{calc}
\usepackage{fancyhdr}
\usepackage{array}
\newcolumntype{P}[1]{>{\centering\arraybackslash}p{#1}}

\usepackage{soul} 

\newcommand{\dz}{\delta{z}}
\newcommand{\dt}{\delta{t}}

\newcommand{\emax}{\epsilon_{\rm max}}

\newcommand{\dd}{{\rm d}}

\usepackage{color}
\usepackage[normalem]{ulem}
\usepackage{xcolor,calc}
\hypersetup{linkcolor = black, citecolor = black, urlcolor = blue}

\begin{document}

\title{Optimizing spin-based terahertz emission from magnetic heterostructures}
\author{Francesco Foggetti}%
\email{francesco.foggetti@physics.uu.se}
\author{Francesco Cosco}
\thanks{Present address: VTT Technical Research Center, Espoo, Finland}
\author{Peter M. Oppeneer}%
\affiliation{Department of Physics and Astronomy, P.\ O.\ Box 516, Uppsala University, SE-75120 Uppsala, Sweden}
\author{Henri Jaffr{\`e}s} 
\affiliation{Laboratoire Albert Fert, CNRS, Thales, Universit{\'e} Paris-Saclay, F-91767 Palaiseau, France}
\author{Niloufar Nilforoushan, Juliette Mangeney, and Sukhdeep Dhillon} 
\affiliation{Laboratoire de Physique de l’Ecole Normale Sup{\'e}rieure,
ENS, Université PSL, CNRS, Sorbonne Universit{\'e},
Université Paris Cité, F-75005 Paris, France}

\date{\today}

\begin{abstract}
    Terahertz radiation pulses can be generated efficiently through femtosecond laser excitation of a ferromagnetic$/$nonmagnetic heterostructure, wherein an ultrafast laser-induced spin current results in an electromagnetic THz pulse due to spin-charge conversion. 
    It is, however, still poorly understood
    how the THz emission amplitude and its bandwidth in the frequency regime can be optimized. Here, we perform a systematic analysis of the THz emission from various magnetic heterostructures. The dynamics of the spin current is described by the semiclassical, superdiffusive spin-transport model and 
    the energy dependence of the spin Hall effect of hot electrons is taken into account, leading to emission profiles for Co(2 nm)/Pt(4 nm) bilayer in good agreement with experiment.
    To identify the optimal conditions for THz emission, we study the properties of the emitted THz wave profile by systematically varying the layer thicknesses of metallic bilayers,
    their interfacial spin-current transmission properties, their materials' dependence, and influence of  the pump laser-pulse width, allowing us to give optimization guidelines. 
    We find that thin nonmagnetic layer thicknesses of $5-6$ nm provide the largest bandwidth in the case of Co/Pt and that the peak frequency of the THz emission depends only on the geometry of the emitter and not on the laser pulse width. The THz bandwidth is conversely found to depend on several factors such as exciting laser pulse width, layers' thicknesses, and interface transmission-reflection properties, with the limitation that an increase in the bandwidth by tuning the interface properties comes with a trade-off in the energy efficiency of the emitter. 
    Lastly, we propose a double pulse excitation protocol of a trilayer system that could provide broadband THz emission with a large bandwidth. Our results contribute to establishing guidelines for optimizing spintronic THz generation.
\end{abstract}
\maketitle

\section{Introduction}

\begin{figure}[!bt]
\begin {center}
   \includegraphics[width=\columnwidth]{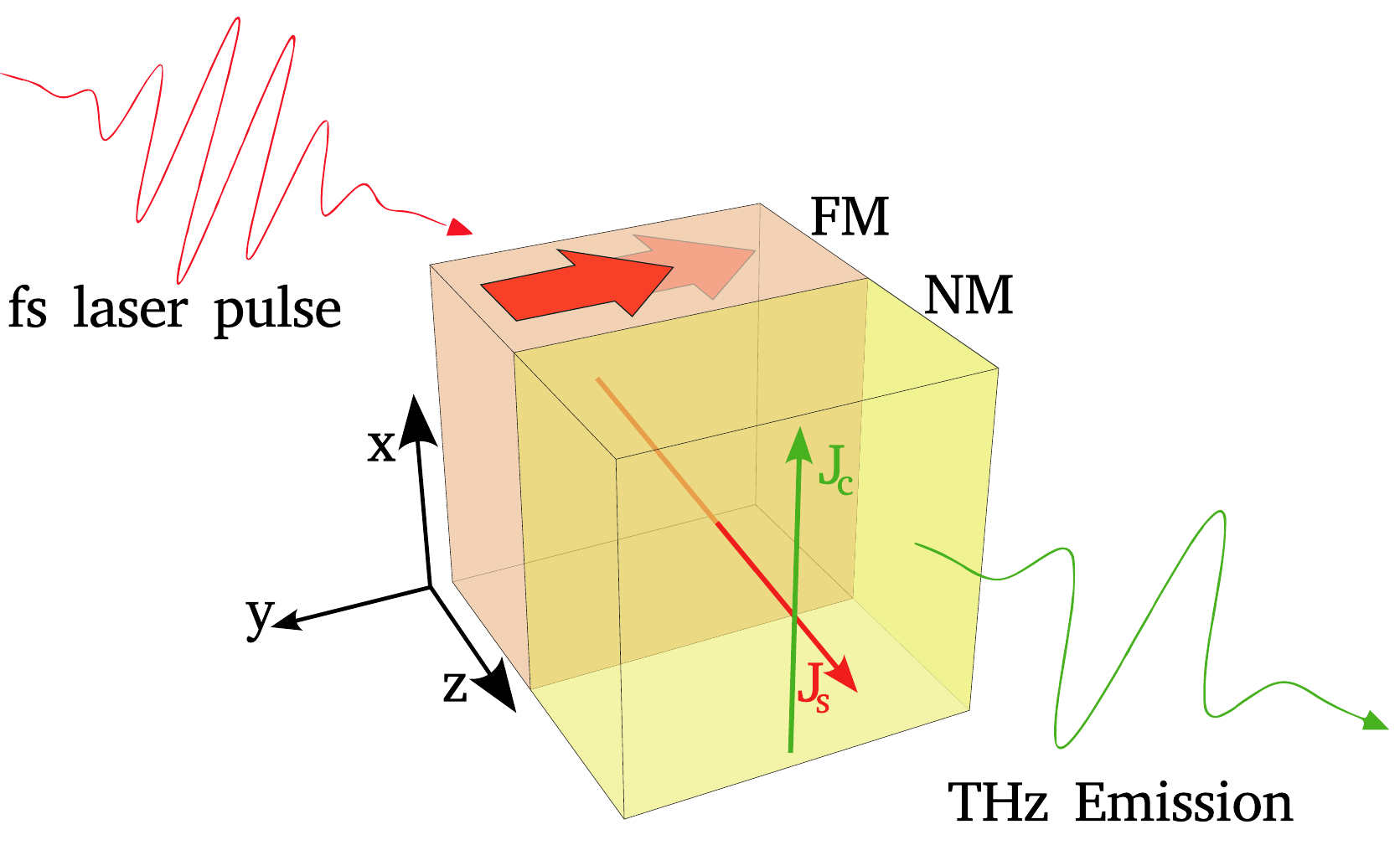} 
\end{center}
\caption{Pictorial representation of a spin-based THz emitter. The fs laser pulse excites  the ferromagnetic (FM) layer, {reducing the magnetization and} generating a {longitudinal} spin current, which is converted to a charge current in the nonmagnetic (NM) layer via the inverse spin Hall effect. The resulting time-varying {transverse} charge current is the source for the THz pulse emission.}
\label {fig1:sketch}
\end{figure}

{The terahertz frequency band is positioned between the gigahertz frequency of electronics and the infrared frequency range of optics \cite{Lee2009,Dexheimer2017,Mittleman2017}. THz radiation  was previously difficult to generate, but in the last decades various THz sources have been developed \cite{Ferguson2002,Tonouchi2007,Dhillon2017,Papaioannou2020,Bull2021}. The most widely used THz emitters, based on THz emission from nonlinear semiconductor crystals via optical rectification or photoconductive antennas \cite{Lee2009,Dhillon2017,Nahata1996,Burford2017,Ferguson2002},  suffer from limited bandwidth capabilities. 

Spin-based THz emitters, discovered a decade ago, provide a technology framework that can overcome this limitation \cite{Seifert2016,Chen2019,Adam2019,Wu_principles_2021}. The basic principle of this technology involves exciting the magnetization of a nanometer-thick heterostructure via a femtosecond laser pulse which triggers an ultrafast demagnetization process of the ferromagnetic (FM) layer, as depicted in Fig.\ \ref{fig1:sketch}. The demagnetization is associated with the generation of a spin current which is injected into a nonmagnetic (NM) metal. In the NM layer the spin current burst is converted in a transverse charge current through spin-to-charge conversion processes that lead to the efficient generation of a THz radiation pulse \cite{Kampfrath2013,Seifert2016,seifert2022,panahi2020,yang2016,Dang2020AppPhys}. The generated short {transverse} charge current causes emission of electromagnetic radiation in the THz regime through electric dipole emission \cite{Kampfrath2013,Zhang2020NatComm}. The spin-to-charge conversion in the NM material, which is typically a heavy metal, occurs through the inverse spin Hall effect (ISHE) \cite{Dyakonov1971,Hirsch1999}, which is a relativistic effect, caused by the spin-orbit interaction \cite{Sinova2015}.
A further important role in the spin-to-charge conversion is played by the inverse Rashba-Edelstein effect (REE) \cite{Bychkov1984, Edelstein1990}, in particular for ultrathin systems where the contribution of Rashba interfaces to THz emission can be comparable or greater than the contribution of ISHE which is a bulk effect \cite{Jungfleisch2018,Zhou2018,Rongione2023}.}

{Although the basic principle of the THz emission is mostly accepted, there remain several open questions. One of these is the proportionality of the THz electric field to the generating source. Different proportionality behaviors have been proposed, such as $E \propto J_c$, or $ \partial J_c/ \partial t$, or $\partial^2 M/ \partial t^2$, and $\partial M/ \partial t$, where {$J_c$ is the charge current and} $M$ is the magnetization of the FM layer, see Refs.\ 
\cite{Beaurepaire2004,Hilton2006,Seifert2016,Huisman2016NatNano,Nenno2019,Pettine2023,Beens2020,Lichtenberg2022,Schmidt2023,Rouzegar2022,Kefayati2024} and references therein. A further important question is, what properties of the metallic bilayer determine the THz emission, and how these properties can be tuned to maximize THz emission signal.}

In this work we address the latter question, through theoretical modeling of the THz pulse generation and the THz wave shape. Our main aim is to understand the factors that influence the THz emission and to provide guidelines for achieving the  maximal THz electric field at a peak frequency $\nu_{\rm max}$ as well as generating a THz emission spectrum that has a high bandwidth.
This kind of modeling is relevant, as control over the THz frequency band of the electromagnetic spectrum is generally believed to promise rich opportunities for the technological advancement of many fields such as spectral imaging, industrial quality control, high-speed electronics, medical diagnostics, and telecommunications \cite  {Tonouchi2007,song2015handbook,Mittleman2017,Chen2020,Lee2009,Dexheimer2017}.

Here,
we systematically study the THz emission caused by superdiffusive spin currents that are generated by pulsed laser excitation in the FM layer and that rapidly penetrate the NM layer (Fig.\ \ref{fig1:sketch})
\cite{Battiato2012,Kampfrath2013,Eschenlohr2013,Melnikov2011}. 
In Sec.\ \ref{sec:methodology}, we introduce the theoretical methodology on which we base our analysis, the superdiffusion  spin-transport model \cite{Battiato2010,Battiato2012,Eschenlohr2013,Balaz2018,Lu2021,Balaz2023}. This model  provides an accurate description of ultrafast demagnetization processes in magnetic heterostructures \cite {Kampfrath2013,Rudolf2012,Hofherr2017,Ritzmann2020:PRB,Gupta2023}.
We then describe how to apply the formalism to obtain the laser-induced spin currents responsible for the THz emission. 

In Sec.~\ref{sec:Results} we report the main results of this paper. First, we start with comparing our model simulations with experimental THz emission data of a Co(2 nm)/Pt(4 nm) bilayer in Sec.~\ref{sec:comparison}, resulting in a very good agreement. Subsequently, we adopt the bilayer geometry of the FM/NM type in the following parts of Sec.~\ref{sec:Results}. We show how the femtosecond laser pulse features, such as pulse duration, influence the THz emission. Then, we characterize the THz spectrum according to the characteristics of the sample, such as the relative thickness of the layers (in the {Appendix} \ref{app:bilayers} we report results for different FM materials). Furthermore, we study the role of interface properties in the THz emission, by varying the reflection properties of the emitter at the central interface, i.e.\ between the FM and NM layer, and at the outer interfaces, between the emitter and outside.
Finally, in Sec.\ \ref{sec:2pulses} we propose a novel kind of setup, a FM/NM/FM trilayer which is excited from both the FM sides by two separate laser pulses with a well defined time delay, which causes interference of the spin currents in the emitter and results in a larger bandwidth profile.

\section{Methodology}
\label{sec:methodology}
\label{sec:SuperDiff}

The superdiffusive spin transport model~\citep{Battiato2010,Battiato2012,Balaz2018,Balaz2023} has been developed to describe in a simple and intuitive fashion the ultrafast laser-induced demagnetization in layered heterostructures of the type  {FM/NM}. In the key idea of the model, a femtosecond laser pulse exciting a ferromagnetic material creates energetic spin-polarized electrons that have anomalous transport characteristics. Hot spin majority and minority electrons have different spectral 
 properties, {specifically,}  
 they move in the sample at different speeds {and have} different lifetimes {\cite{Zhukov2005,Zhukov2006,Nechaev2010}}. These differences give rise to a spin-polarized diffusion which is responsible for the demagnetization of the ferromagnetic layer as the induced, longitudinal spin current leaves the magnetic layer and is injected in the nonmagnetic material, see Fig.\ \ref{fig1:sketch}.

In our study, a laser pulse initiates the electronic transport by exciting 
electrons with energies at and below the Fermi level -- typically occupying strongly
$d$-hybridized and relatively immobile states -- into higher energy bands of $sp$ 
character. These $sp$ electrons are characterized by higher velocities, hence, they behave as 
itinerant particles moving through the sample~\citep{Battiato2014}. We note that an alternative approach to describe the collision-rich motion of excited hot electrons in metallic layers has been developed on the basis of Boltzmann transport theory \cite{Nenno2016,Nenno2018}.

The superdiffusive transport model describes two spin channels, 
tagged as $\sigma\in\{\uparrow,\downarrow\}$,  for the electronic transport. 
Each spin channel is characterized by the hot electron velocities,  $v_{\sigma}(\epsilon)$, and lifetimes,  $\tau_{\sigma}(\epsilon,z)$,  which depend on the electron energy $\epsilon$,
and, in magnetic materials, on the spin $\sigma$ as well. The  lifetime $\tau$ is a $z$-dependent quantity due to the layered structure of the emitter,
wherein different materials along $z$ have different lifetimes.
Because of the distinct transport properties of the two spin channels, the current of hot electrons {becomes}
spin polarized in the magnetic layer.
In addition, in the case of multilayers,  spin filtering
via multiple spin-dependent transmissions and reflections at the interfaces further 
contributes to the spin polarization of the current~\citep{Battiato2014}.

The outflow of the spin-polarized current $J_s$ 
initiated by a laser pulse results in a loss of local magnetic
momentum and {corresponding} demagnetization $ dM/dt$.  Due to the high velocities of hot electrons,  
the demagnetization happens {typically} on the timescale of {a few hundred} femtoseconds~\cite{Battiato2010}. 
{In the superdiffusive model} the hot electrons propagate {initially ballistically} {in approximately the first 100\,fs.} However, due to the relaxation processes caused by electron scattering {and thermalization, the hot electron transport} continuously changes its character.  
In about $500\,{\rm fs}$ up to $1\,{\rm ps}$ it develops into diffusive {transport}.  
In the time interval up to 500 fs,  
the transport proceeds in the superdiffusive regime~\citep{Battiato2012}.

Let us describe briefly the main features of the superdiffusive spin-dependent
transport model,
which revolves around the equation  of motion for the  hot electron density $n_\sigma (\epsilon,z,t)$, with spin $\sigma$, energy $\epsilon$, and position $z$,
\begin{equation}
\frac{\partial n_\sigma (\epsilon,z,t)} {\partial t}+ \frac {n_\sigma (\epsilon,z,t)}{\tau_\sigma (\epsilon,z)}=
 \left (  -\frac {\partial  } {\partial z}  \hat \phi+ \hat I\right) S_\sigma^{\mathrm {eff}} (\epsilon,z,t),
\label{eq:eom}
\end {equation}
where $\hat \phi$ is the flux operator which contains the dependence from the electrons' velocities 
and describes interlayer  transmissions and reflections.  Furthermore,  $S_\sigma^{\mathrm {eff}} (\epsilon,z,t)$
is an effective source term  describing the laser induced excitation of spin-polarized hot electrons
and scattering events within the material. 
{It consists of two contributions, 
\begin{equation}
S_\sigma^{\mathrm {eff}}(\epsilon,z,t)=S_{\sigma}^{\mathrm {ext}}(\epsilon,z,t)+S_{\sigma}^{{\rm p}}(\epsilon,z,t)\,,
\label{Eq:gen_source}
\end{equation}
where $S_\sigma^{\mathrm {ext}}(\epsilon,z,t)$ describes the excitation of hot electrons by the laser pulse, and $S_{\sigma}^{{\rm p}}(\epsilon , z,t)$ is the term describing the effects of scattering, mostly caused by electron-electron interactions, calculated as
\begin{equation}
\begin{split}
S_{\sigma}^{{\rm p}}(\epsilon,z,t+\delta{t})= & \sum_{\sigma'}\int_{\epsilon_{\mathrm {Fermi}}}^{\emax}\dd\epsilon'\;n_{\sigma'}(\epsilon',z,t)\times\\
p_{\sigma^{\prime},\sigma} &(\epsilon',\epsilon,z,t)  \left(1-e^{-\delta{t}/\tau_{\sigma'}(\epsilon',z,t)}\right) .
\end{split}
\label{Eq:Sp_def}
\end{equation}
$p_{\sigma',\sigma}(\epsilon',\epsilon,z,t)$ is
the probability that an electron at energy level $\epsilon'$, between the Fermi energy $\epsilon_{\mathrm {Fermi}}$ and the energy cut-off 
$\epsilon_{\mathrm {max}}$, and
spin $\sigma'$ will move to energy level $\epsilon$ with spin $\sigma$
in the next time step, $t+\delta{t}$. 
Eq.~(\ref{Eq:Sp_def}) includes both the contributions from the scattered hot electrons, formally treated as newly excited,
and the ones actually excited as the result of the scattering.}
For practical calculations we adopt the discretized version of the formalism, described in Ref.~\cite{Battiato2014}, 
with the space divided into computational cells of width $\delta z$. Time and energy are also sampled in finite steps 
of $\delta t$ and $\delta \epsilon$, respectively.  In this {article}, we use the spatial discretization step $\dz = 1$~nm,
time step $\dt = 1$~fs, and energy step $\delta\epsilon =0.125$~eV.

When treating the multilayers, one should include the possibility of scattering at the interlayer interfaces. 
{This is done by employing positions on the left and right-hand side of the interface,} 
 $z_i^{\pm}=z_i\pm\dz/2$,  for an interface 
centered at $z_i$, and computing the fluxes  $\overrightarrow{\Phi}_{\sigma}(z_i^{\pm},t)$ 
and $\overleftarrow{\Phi}_{\sigma}(z_i^{\pm},t)$ of right and left
moving particles, respectively,  with spin $\sigma$ that go  through the interface $z_i^{\pm}$ at time $t$. 
Each interface is characterized by spin and energy-dependent transmission or reflection coefficients for electrons moving to the right,
or to the left, that can be obtained from first-principles calculations \cite{Battiato2014,Dang2020AppPhys,Lu:PRB2020,Balaz2023}.

The solution of the discretized version of Eq.~\eqref{eq:eom} gives automatically access to the spin current density,  defined as the difference between spin-up ($\uparrow$) and spin-down ($\downarrow$) electron flux densities  flowing in and out ($\rightarrow$,$\leftarrow$) at each point in space, i.e.,
\begin {equation}
J_s(z,t)= \frac {\hbar} {2} \left [\overrightarrow{\Phi}_\uparrow (z,t)-\overleftarrow{\Phi}_\uparrow (z,t)-\overrightarrow{\Phi}_\downarrow (z,t)+\overleftarrow{\Phi}_\downarrow (z,t)   \right].
\end {equation}
The spin current is converted via ISHE into a transient charge current as
$\mathbf{J}_c(z,t) =  \frac {2 e}{\hbar}\theta_{\text{SH}} \mathbf{J}_s(z,t)\times \mathbf{M}/|\mathbf{M}|$,
where $\theta_{\text{SH}}$  is the spin Hall angle, and represents the efficiency of the spin-to-charge conversion process, and $\mathbf{M}$ is the magnetization of the FM layer. 

{An important remark has to be made at this point, namely, that} the spin Hall angle is energy dependent. {It is defined as} $\theta_{\text{SH}} = \sigma_{\text{SH}}/\sigma$, where $\sigma_{\text{SH}}$ is the spin Hall conductivity and $\sigma$ is the usual conductivity of the material \cite{Stamm2017}. 
In particular, the spin Hall conductivity will depend on the energy of hot electrons \cite{Salemi2022}. Due to the large spin-orbit interaction in Pt, a sizeable
spin-charge conversion is obtained in spintronic emitters,  making Pt a standard choice for the heavy-metal to realize
high-amplitude spintronic emitters \cite{Seifert2016,Wu2017,Seifert2018,Zhang_2018, Dang2020AppPhys,Adam2019,Papaioannou2020}.
The energy-dependent spin Hall conductivity {of Pt} is large for hot electron energies close to the Fermi energy, but drops significantly for energies higher than 0.5~eV. The origin of this energy dependence is the 5$d$ band structure of Pt, in which the top of the 5$d$ band lies about 0.5~eV above the Fermi energy. Hot electrons with higher energies are injected into the Pt $sp$ band, but the predicted spin Hall conductance for these electrons is much smaller. This means that hot electrons injected in the Pt layer will contribute differently to the emission, in particular, electrons with higher energy first need to lose energy, i.e., by scattering and decaying into lower energy levels, in order to give a significant contribution to the THz emission.
As outlined recently, the contributions to the spin-to-charge conversion from the different hot-electron energy levels can be incorporated in the superdiffusive transport, see Ref.\ \cite{Foggetti2025} for details. {This renders the spin and charge currents to be disproportional.} In the following simulations of spintronic THz emission, we include the energy dependence of the ISHE when we numerically compute the charge current.

Lastly, the time-varying charge current is responsible for the emitted THz signal. It has been discussed recently whether the emitted THz electric field would be proportional to the charge current $J_c$ \cite{Seifert2016,Huisman2015PRB,Nenno2019} or the time derivative of the charge current, $\partial J_c/ \partial t$ 
\cite{Kefayati2024,Pettine2023,Varela2024}. Although according to the Maxwell equations the electric amplitude in the far field should be proportional to $\partial J_c/ \partial t$ \cite{Jefimenko1992}, the parabolic mirror used in the focusing optics of the THz detection leads to an effective time integration so that, in the frequency domain, one obtains \cite{Foggetti2025} 
\begin{equation}
    E(\nu)\propto\int dz \, J_c(z,\nu). 
\end{equation} 
 This proportionality will be employed in the following simulations for the THz electric field from spintronic emitters.

\section{Results}
\label{sec:Results}

\subsection{Comparison with experiment}
\label{sec:comparison}

{Before delving into the possibilities to optimize the spintronic THz emission from spintronic heterostructures, we first compare the   emission calculated with our methodology with measured THz emission, to examine how well the theoretical model is able to describe the experiment.}
 Specifically, we consider a realistic Co/Pt heterostructure and we compare our modeling results with the {measured} THz spectrum emitted from a Co(2)/Pt(4) bilayer.
 Here and in the following, the numbers in parenthesis refer to the layer width in nm.  A laser pulse of wavelength 1030 nm (1.2~eV) and {width} $\tau_\mathrm{pulse}=$~23~fs from an amplified laser system excites a hot spin-polarized current in the Co film and injects it in the Pt, where it is converted into a charge current that results in the THz pulsed emission. A THz-time domain spectroscopy setup is used to detect the amplitude and phase of the emitted THz pulse. For ultra-broadband detection a 20 $\mu$m thick ZnTe crystal is employed for electro-optic sampling of the generated THz pulse  
 \cite{Hale14,Nilforoushan2022}.
 
To compute the THz emission we employ the superdiffusion model in combination with spin-dependent electron lifetimes and velocities as given by first-principles calculations \cite{Zhukov2006,Nechaev2010}.
{Here, the external boundaries of the heterostructure, representing the capping material and substrate that enclose the emitter, are considered completely reflective (R$\,=1$), {because the measured sample was capped with AlO$_x$ and prepared on a sapphire substrate}. In order to estimate the reflection coefficients at the Co/Pt interface, we use the reflection coefficients reported in Ref.\ \cite{Lu:PRB2020}, where, however, only the reflection coefficients for Fe and Ni are presented. Thus, we decide to approximate the coefficients for Co as an average of these two.}
{Furthermore, we take the energy dependence of the spin Hall conductivity and the response function of the $20$-$\mu$m ZnTe detector crystal into account (see \cite{Foggetti2025}) as well as the integrating influence of the focusing mirror.}

\begin{figure}[tb!]
    \centering
   \includegraphics[width=\linewidth]{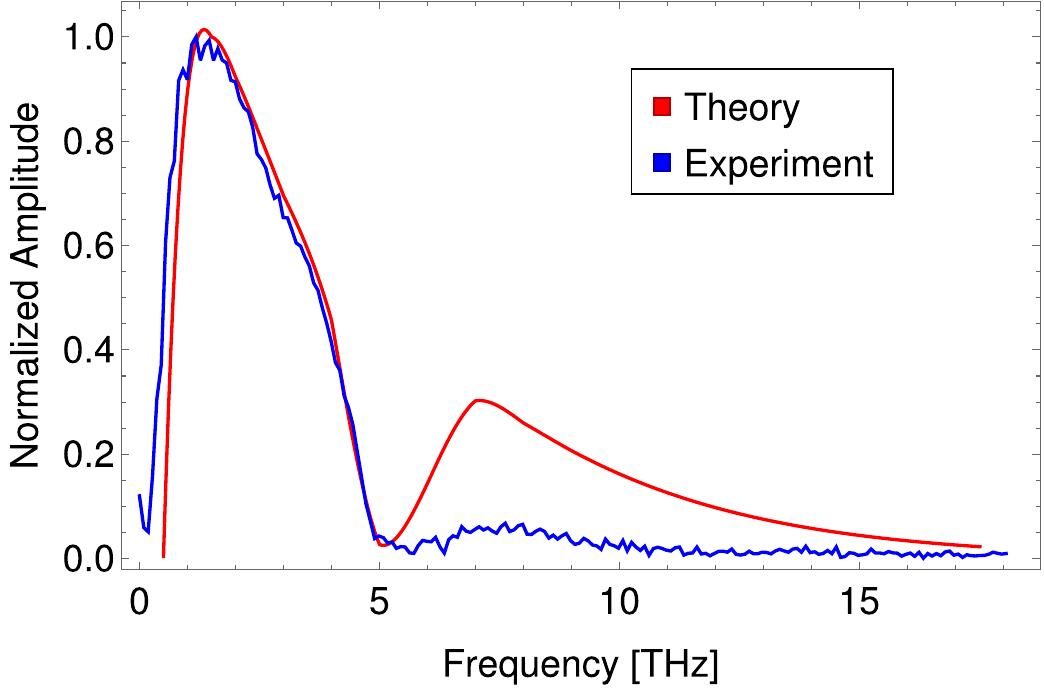}
    \caption{
Comparison of our modeling (red) and experiment (blue) for the THz amplitude spectrum of a Co(2 nm)/Pt(4 nm) emitter. The response function of the ZnTe detector is taken into account and is responsible for the dip at 5 THz. 
The THz emission amplitude $E$ is normalized to its  maximum value. }
    \label{OPT-fig:emissionZT}
\end{figure}

Figure~\ref{OPT-fig:emissionZT} shows the simulated emission profile compared with the experimental data for a Co(2)/Pt(4) spintronic THz emitter. 

The simulated THz emission spectrum is in good agreement with the experimental data: the {frequency} position of the emission peak is correctly {predicted} 
{and} the bandwidth is {also in good agreement.} 

The secondary peak that appears in Fig.~\ref{OPT-fig:emissionZT} {at $7-8$ THz} is {related to the} 
response function of the {ZnTe} detector, {which has a dip at 5 THz.} 
Such a peak is present, too, in the experimental data, although smaller in amplitude. 

{The good agreement between {the first-principles-based} and measured THz emission spectra ascertains that our modeling provides an adequate description of the spintronic THz emission process. Having established this, we consider in the following how the THz emission can be optimized.} 
{If not stated otherwise, we use completely reflective boundaries at the outer interfaces of the emitter.  
In the following, we focus on how the THz emission from a typical Co/Pt bilayer can be optimized.
In Appendix~\ref{app:bilayers} we provide calculated results for Fe/Pt and Ni/Pt emitters, using the internal FM/NM reflection coefficients reported in Ref.\ \cite{Lu:PRB2020}.}

\subsection{Simulated results for FM/NM bilayers}

{The FM/NM bilayer system} is considered {as} the {fundamental} element of THz spin-based technologies \cite{Kampfrath2013,Papaioannou2018,Wu2017,Li2019,Dang2020AppPhys,Bull2021,Feng2021}. 
In what follows, we assume that the laser always excites the sample from the ferromagnetic side, as depicted in Fig.\ \ref{fig1:sketch}. 

The aspects of the setup that can influence the shape and characteristics of the spin current and, therefore, the emitted THz emission, can be divided into two categories: passive and active characteristics.
In the passive category, we place those elements which we cannot vary from experiment to experiment using the same sample. These are the physical properties of the experimental setup, i.e., the specific materials composing the multi-layered heterostructure, the quality of the interfaces, the number of layers and their {thicknesses}.
In the active category, we place those elements that are not intrinsic to the setup and can be engineered at will. These include the properties of the laser pulse exciting the sample, which can be the pulse length, the number of pulses, and the intensity. 
{In the {following} considerations, we will not include the influence of the electro-optic detector crystal. Such detector crystal is always present in the measurements, but different materials are being used (GaAs, ZnTe, etc.) and with varying thicknesses, which makes the recorded THz spectra setup dependent. Our focus is thus on the THz emission before it is measured with the electro-optic crystal.}

To characterize the property of the emitted THz signal, we consider two figures of merit extracted from the frequency profile of the emitted electric field $E(\nu )$. The first one is the peak frequency $\nu_{\mathrm {max}}$, which is the frequency of the maximum of $E(\nu)$. The second one is the signal bandwidth, which we define as the frequency interval in which the signal is reduced by one order of magnitude with respect to the maximum $E(\nu_{\mathrm  {max}})$. 
A further quantity that can serve as an indicator is the total integrated power of the emitted radiation, 
\begin{equation}
    P \propto \int_0^{\infty} d \nu \, | E (\nu ) |^2 ,
\end{equation}
which gives a {reasonable} measure for the {amplitude of the generated THz radiation.} 

\subsection{Influence of laser-pulse width}
\label{sec:tau_pulses}

\begin{figure}[!h]
\begin{center}
\includegraphics[width=\columnwidth]{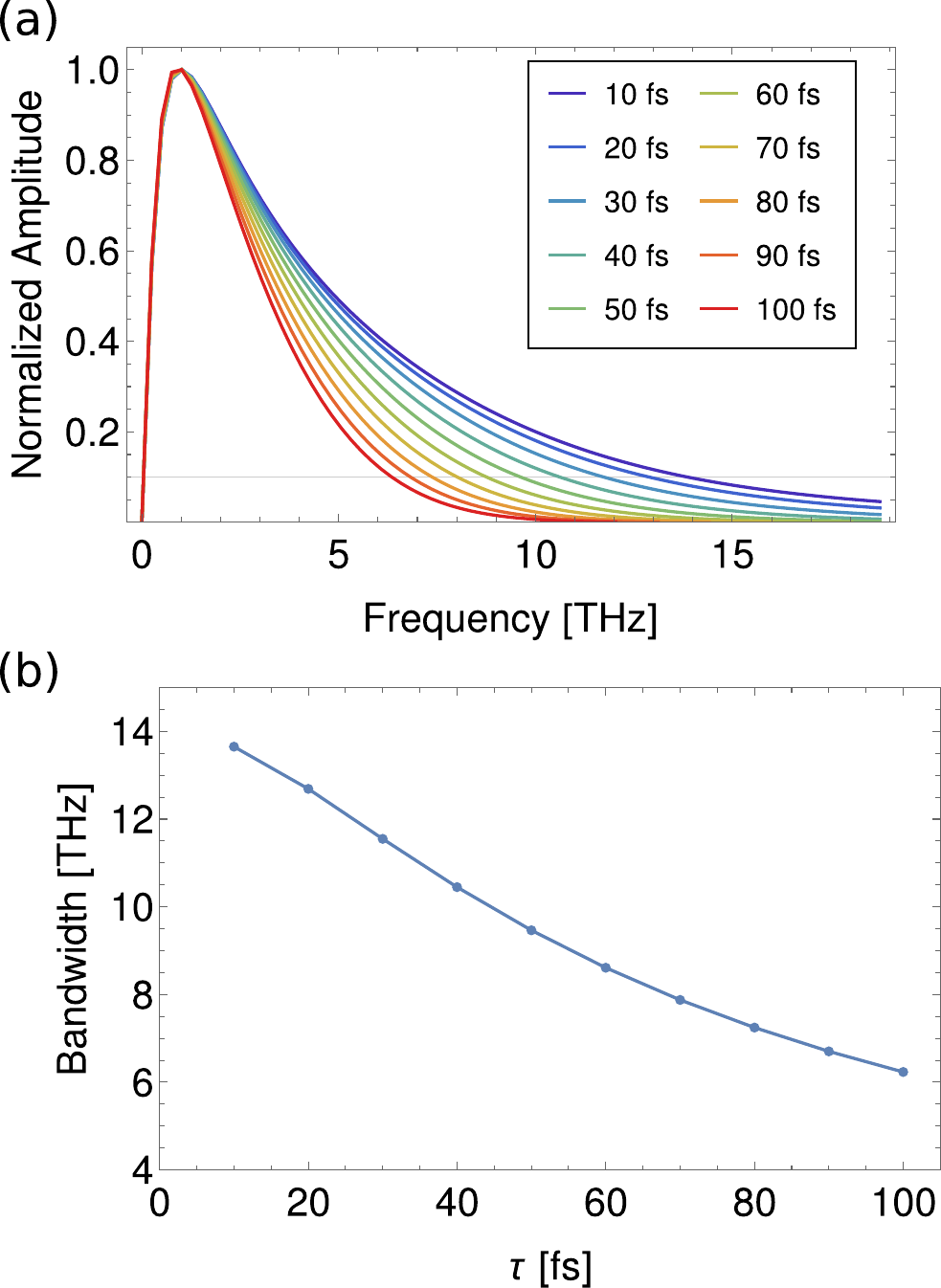}
\end{center}
\vspace*{-0.3cm}
\caption{
Results of THz emission calculations. (a) THz emission spectrum of a Co(2)/Pt(4) bilayer computed for different pump durations. (b) Computed THz bandwidth as a function of the femtosecond pump pulse duration. 
}
\label{fig:taus}
\end{figure}

We start our analysis by exploring the effects of the laser-pulse properties on the THz spectrum. Figure \ref{fig:taus}(a) shows how the {simulated} signal bandwidth {of a Co(2)/Pt(4) bilayer} varies with the {pump}-pulse length, {defined as the full width at half maximum (FWHM) of the pulse}. 
When reducing the {pump}-pulse duration, we observe an increase of the bandwidth in Fig.\ \ref{fig:taus}(b). {Such bandwidth increase was noted previously  by Nenno \textit{et al.} \cite{Nenno2019}.}
This {effect} is to be expected since the emission dynamics is {triggered} by the laser pulse, i.e.,\ the {pump}-pulse duration {sets} the time-scale during which electrons are excited to above the Fermi energy \cite{Nenno2019}.   
For a short {pump} pulse, the emission is strongly dominated by the fast dynamics of the {excited} hot electrons. {These start to} act as a signal source as soon as they reach the nonmagnetic layer. {Therefore}, the resulting THz spectrum is {more strongly} characterized by high-frequency components. 

Interestingly, we observe that the peak frequency of the signal is nearly not affected by the change in pulse duration. As we will show in the following sections, it appears instead that the peak frequency of the signal changes when the geometry of the sample, i.e., layers' thickness  and interface properties, is altered. 
{The minor dependence of the peak frequency on the pulse duration is consistent with the expectation that during long pulses hot spin-polarized electrons are injected into the NM material for a longer time, thus resulting in a spectrum peaked at a lower frequency.}

{Finally, we} mention that no effect is instead observed on the bandwidth {when} the intensity of the laser pulse is changed. The only observed effect is a change in the height of the emission peak, but the {time profile} of the {spin current} remains unaffected.
{This holds within moderate variations of the intensity, {in the linear regime of excitation (not complete demagnetization),} where the generated spin current scales linearly with the adsorbed laser fluence.

\subsection{Influence of layer thicknesses}
\label{sec:layer_size}
\begin{figure}[!h]
\begin {center}
\includegraphics[width=.75\linewidth]{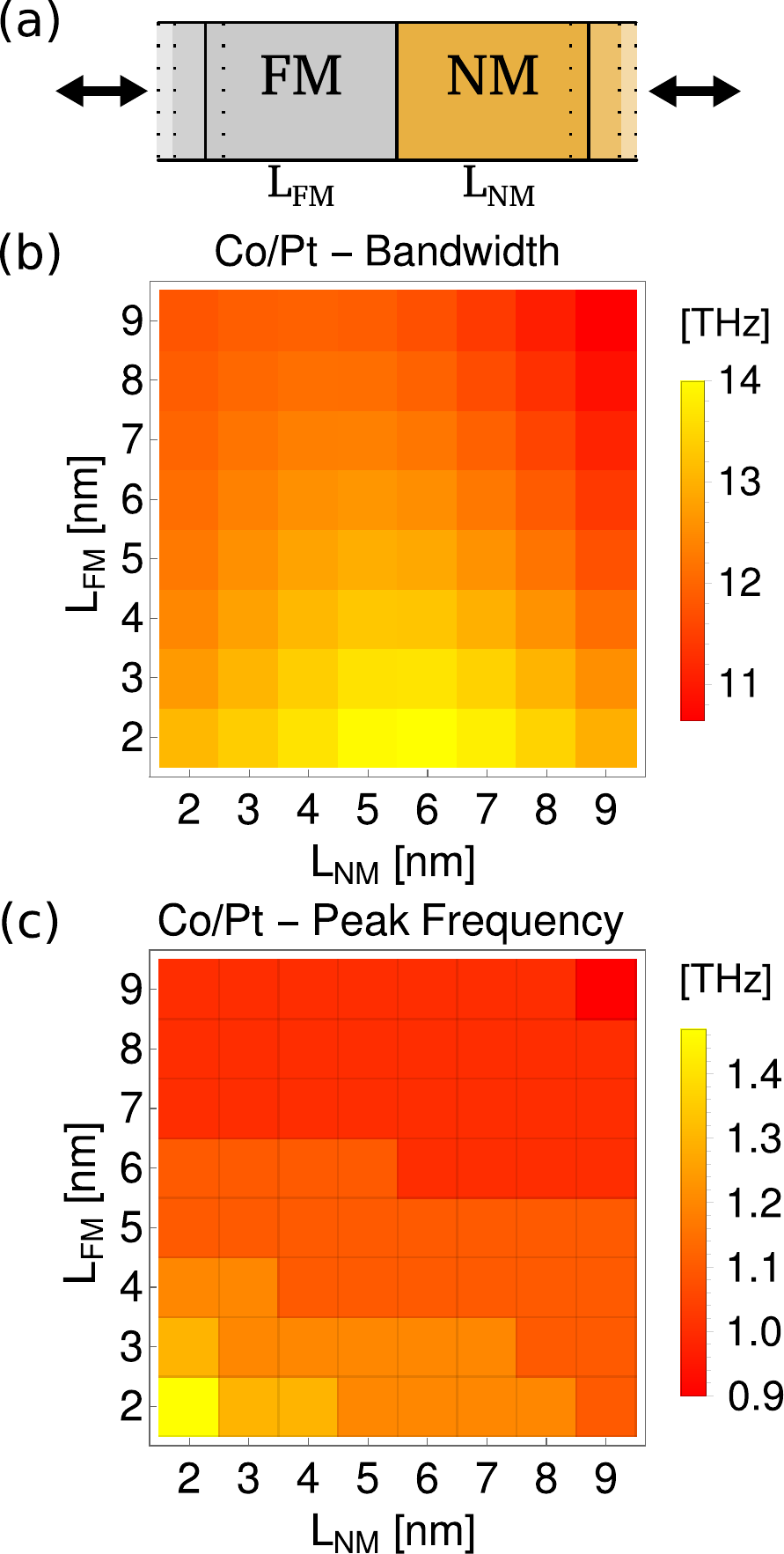}
\end{center}
\vspace*{-0.4cm}
\caption {(a) Schematic representation of the {variation of} 
layer thicknesses of the spintronic THz emitter. The thicknesses ${\text L}_{\text{FM}}$ of the FM and ${\text L}_{\text{NM}}$ of the NM layer are varied independently to study their influence on the THz properties of the system. (b) Simulated THz bandwidth and (c) peak frequency of a Co($\mathrm{L_{FM}}$)/Pt($\mathrm{L_{NM}}$) emitter as a function of the ferromagnetic and nonmagnetic layer thicknesses, $\mathrm {L_{FM}}$ and $\mathrm {L_{NM}}$, respectively. The grid in the peak frequency plot has been added for better visualization.}
\label{fig:size}
\end{figure}

We continue our analysis by exploring the effect of the system size on the THz emission spectrum. In the following calculations, we {set} the pump-pulse duration to be $\tau_{\mathrm {pulse}}=10$ fs.}
Figures~\ref{fig:size}(b) and  \ref{fig:size}(c) show the bandwidth and the peak frequency, {respectively,} {computed} as a function of the thickness of the FM layer, $\mathrm {L_{FM}}$, and the thickness of the NM material, $\mathrm {L_{NM}}$, {schematically shown in Fig.\ \ref{fig:size}(a)} for a Co/Pt bilayer.  We consider a minimum thickness for both layers of 2~nm, as calculations with thinner layers in presence of reflections at the central FM/NM interface were computationally prohibitive. Further calculations including thinner layers performed neglecting the central reflections suggest no change in the trend of Figure~\ref{fig:size}.
We {readily observe} how larger layer thicknesses reduce the signal's bandwidth and, at the same time, {modify} the peak frequency. 
These findings can be explained by examining the dynamics of the spin current in the superdiffusive model. In the thinner samples, the hot electron wave generated by the fs-laser pulse travels back and forth through the layers before the {hot} electrons' energies completely decay {due to scattering.} Consequently, the current profile {arises} from spin-polarized electrons coming from both sides, destructively interfering. This means that the spin current has a short temporal width and, therefore, a large bandwidth. In the thicker sample, the spin current profile is reduced only by the natural decay time, thus appearing as a pulse with a larger {spread} 
in time and {hence} a shorter bandwidth. 

{Optimal layer thicknesses for a large bandwidth are found to be $\mathrm{L_{FM}} \approx 2 - 3$ nm and $\mathrm{L_{NM}} \approx 5 - 6 $ nm.} 
If the Pt thickness is shorter than the Pt diffusion length {of about} 3~nm \cite{Dang2020AppPhys}, {hot} electrons {injected into} Pt can easily reach the FM layer {again}. Here the electrons can {be adsorbed in the FM layer} {due to} scattering events 
and not be injected into the Pt layer a second time, effectively reducing the emission from {spin-charge conversion in} Pt.

The {computed} variation of the peak frequency {in Fig.\ \ref{fig:size}(c) is relatively small},
{yet it shows a shift of}
{the peak frequency to} {lower frequencies with increased Pt thickness}. 
{The reason for this behavior is that thicker layers of Pt {are favorable for} longer-lived excited electron currents, but longer signals result in shorter bandwidths, which correlate well with smaller peak-frequency values. 
In a thin layer, currents reflect multiple times at the interfaces, interfering destructively and resulting in short-lived signals. In a thicker layer, on the other hand, the current is reflected less often and persists for longer periods of time, comparable to the lifetimes of the excited electrons.}
{This results in a shift to lower peak frequencies for thicker layers.}

{Together,} the competition between the need of a {NM} layer thin enough that favors destructive interference of the currents but not excessively thin that the electrons are easily injected back in the FM layer, selects the optimal thickness for emission that we observe in Fig.~\ref{fig:size}(b),(c),  which is in {overall} agreement with experimental results \cite{Seifert2016, qiu2018layer}.

Another feature that could play a role in the reduction of THz bandwidth from thick Pt layers is the self adsorption in Pt of THz radiation. However, we did not take it into consideration here as the penetration length for THz radiation (of {the order of tens of nm}) is much {larger} than the thicknesses in our simulations.

{With regard to the materials' dependence,} it is worth mentioning that a similar THz emission behavior is obtained, if we use Ni instead of Co for the FM layer, but not for Fe. Details are given in Appendix~\ref{app:bilayers}. This is due to the unique transport properties, including the interactions with the interfaces, of laser-excited electrons in {the} three FM metals which result in a different injection of spin-polarized electrons into the NM layer and different secondary injection of {back-diffused} electrons from the FM into the NM layer. 

\subsection{Interface transmission properties}
\label{sec:interface}
\begin{figure}[th!]
\begin {center}
\includegraphics[width=.8\linewidth]{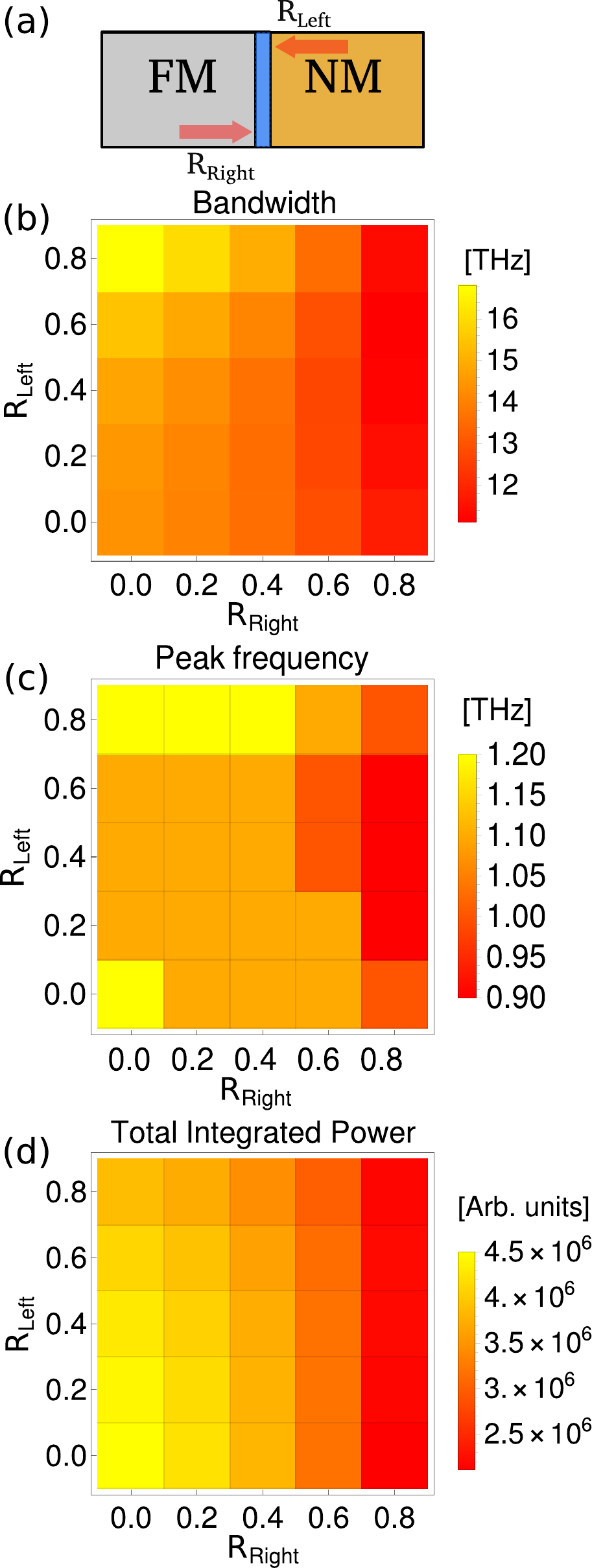}

\end{center}
\vspace*{-0.5cm}
\caption{
(a) Schematic illustration of the interface transmission properties of the spintronic emitter.  
(b) Simulated signal bandwidth, (c) peak frequency, and (d) total integrated THz power as a function of the interface reflection coefficients for left- and right-moving electrons, $\mathrm {R_{Left}}$ and $\mathrm {R_{Right}}$, respectively. The signals are {calculated for a} Co(2)/Pt(4) bilayer.}
\label{fig:filter}
\end{figure}

{Besides the thicknesses} of {the} FM and NM layers, the interface between the FM and the NM layer plays a fundamental role, {too,} in determining the THz {emission.}
The separating region between the two layers {contributes in two different ways}, depending on the direction of the spin current that crosses this region, {see Fig.\ \ref{fig:filter}(a)}.  Considering the polarized spin current coming from the FM side and injected in the NM layer (right-moving electrons), it acts as a filter, reducing the amount of current that contributes to the THz emission. On the other side, 
it acts as a barrier for the {spin} current flowing from the NM part of the emitter (left-moving electrons), preventing it from being diffused again into the FM layer, {thus} trapping the electrons in the NM region, where the spin current is converted in charge current due to ISHE.

In Fig.\ \ref{fig:filter}(b) we show  the computed bandwidth for different values of the interface reflection coefficients {in an Co(2)/Pt(4) bilayer}. For simplicity, we assume the reflection coefficients at the interface to be independent of the electrons' energy, and the interfaces with the outside of the emitter to be perfectly reflecting (R$\,=1$). For comparison, the energy-dependent reflection coefficients are reported in Appendix~\ref{app:ref_coeff}. The case with perfectly adsorbing external interfaces (R$\,=0$) is {discussed} in Appendix~\ref{App:interfaces}.
We study the variation of the bandwidth for different values of the reflection coefficients with respect to right-moving (R$_\mathrm{Right}$) and left-moving (R$_\mathrm{Left}$) electrons. We observe that a maximum bandwidth is obtained for high values of R$_\mathrm{Left}$, meaning that the electrons coming from the NM layer are mostly reflected back and constrained within the NM layer, giving a higher contribution to the emission than if they were injected {back into} the FM layer. 
With respect to the right-moving electrons, the optimal bandwidth is obtained for low reflection coefficients R$_\mathrm{Right}$, meaning that the preferred configuration is one that favors electrons' injection from the FM into the NM layer, giving again a positive contribution to the spin current in the NM layer that further increases the THz emission signal.

It is {furthermore} interesting to analyze how the total integrated power is affected by the changes of the reflection coefficients at the interface. 
We observe in Fig.\ \ref{fig:filter}  
{that} the maximum integrated power is obtained for low values of the R$_{\mathrm{Right}}$ coefficient,
meaning that a more ``transparent'' interface is ideal for a high integrated power, as electrons can freely diffuse into the NM layer and participate in the THz emission. {Comparing with the bandwidth maximum, {occurring for} high values of R$_\mathrm{Left}$ and low values of R$_\mathrm{Right}$, the total integrated power for these reflection coefficients is slightly lower.} 
Finally, {a} minimum integrated power is obtained for high values of R$_\mathrm{Right}$, which means that very few electrons can pass from the FM into the NM layer, so the emitted electric field is at its lowest in this configuration.

\subsection{Boundary properties}
\label{sec:boundary}

\begin{figure}[t]
\begin {center}
\includegraphics[width=.8\linewidth]{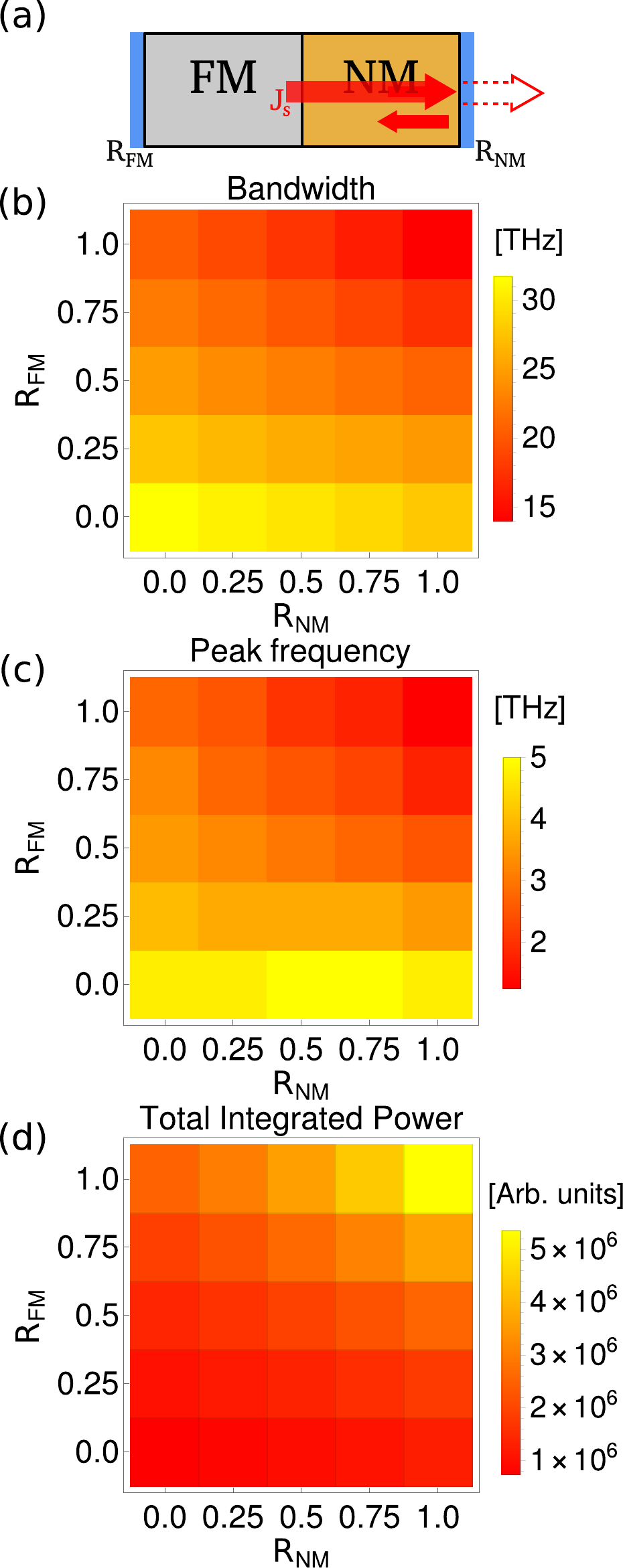}

\end{center}
\vspace*{-0.5cm}
\caption {(a) Schematic illustration of the influence of the outer boundaries of the spintronic THz emitter.
(b) Simulated signal bandwidth, (c) peak frequency, and (d) total integrated THz power as a function of the external boundary reflection coefficients for the ferromagnetic and nonmagnetic layers,  $\mathrm {R_{FM}}$ and $\mathrm {R_{NM}}$, respectively. The signals are simulated numerically {for a}  Co(2)/Pt(4) bilayer. }
\label{fig:boundary}
\end{figure}

Next, we move our analysis to focusing on the role of the outer boundaries of the system. 
In any realistic scenario, the spintronic source will be grown on a substrate or protected with a cap material. Thus, the properties of the boundaries between the bilayer and the external environment, see Fig.\ \ref{fig:boundary}(a), are expected to impact the shape of the THz spectrum. We can distinguish two extreme cases: when the boundary is completely reflective, thus the spin current perfectly bounces off the edge and is injected again into the system, and the case in which no component of the spin current is reflected back into the system and the boundary acts as a so-called \textit{spin sink} \cite{harii_spin_2012,Dang2020AppPhys,hawecker2022spintronic}.

In  Fig.\ \ref{fig:boundary}(b)-(d) we show the computed bandwidth, the peak frequency, and the total integrated power for the Co(2)/Pt(4) bilayer. As in the previous case, the reflection coefficients at the boundary are considered energy independent, {while the reflection coefficients at the inner Co-Pt interface have not been changed.} 
We can distinguish some general trends: the bandwidth {is larger} for lower values of the reflectivity at the boundaries, with a slightly higher {bandwidth} enhancement for low reflection {coefficients} at the FM {outer} boundary; an asymmetry which is related to the nonequal {thicknesses} of the two layers. 
We can understand this behavior of the bandwidth {further} by considering the time scale of the current and the scattered electrons in the two limit cases. 
For completely adsorbing boundaries (R$=0$, {i.e., spin sink}), the timescale related to the injection of the spin current is only determined by the pulse duration,  the hot electrons' speed, and {their} lifetime. In absence of reflections at the boundaries, the spin current generated in the FM layer propagates through the NM layer and is then {evicted} from the system, resulting in the shortest possible current, which translates in the largest bandwidth of the emitted signal. On the other hand, as shown in Fig.~\ref{fig:boundary}(d),  the total {emitted} power is the lowest in this case, meaning that a great amount of {injected} energy is lost due to the {spin current} adsorbing boundaries. 

In the case of completely reflecting boundaries (R$=1$) however, there are two {consecutive} contributions to the signal: the hot electrons that reach the {NM} layer right after being excited, as well as a portion of electrons which first scatter off the outer boundary surface before reaching the NM material and are injected in the NM layer at later times.
This results in a longer {spin-current} dynamics, which translates into a {smaller spectral} bandwidth. 
The integrated power is maximum in this case, up to a factor of five greater than the completely adsorbing boundaries, as there is no loss of {hot} electron energy at the boundaries.  

Moving forward, we observe an interesting pattern in the peak frequency, Fig.~\ref{fig:boundary}(c). The peak frequency  is  larger for lower reflectivity at the FM outer boundary, similar to the trend seen in the bandwidth.
{Relatively large} peak frequencies are still possible in {the} presence of moderate reflection (up to R$_{\text{NM}}=0.5$) at the NM layer. 
The peak frequency reaches its minimum 
when both the external boundaries are completely reflective.
This behavior can be explained using similar arguments as for the signal bandwidth. When the external boundary of the NM material is completely reflective, the generated spin current is long lived, and the resulting THz signal has significant low-frequency contributions. Conversely, in the {spin-sink} case, the currents in the system are quick and short-lived, resulting in signals with more pronounced higher frequency components.

\subsection{Double pump-pulse emitters}
\label{sec:2pulses}
\begin{figure}
    \centering
    \includegraphics[width=.76\linewidth]{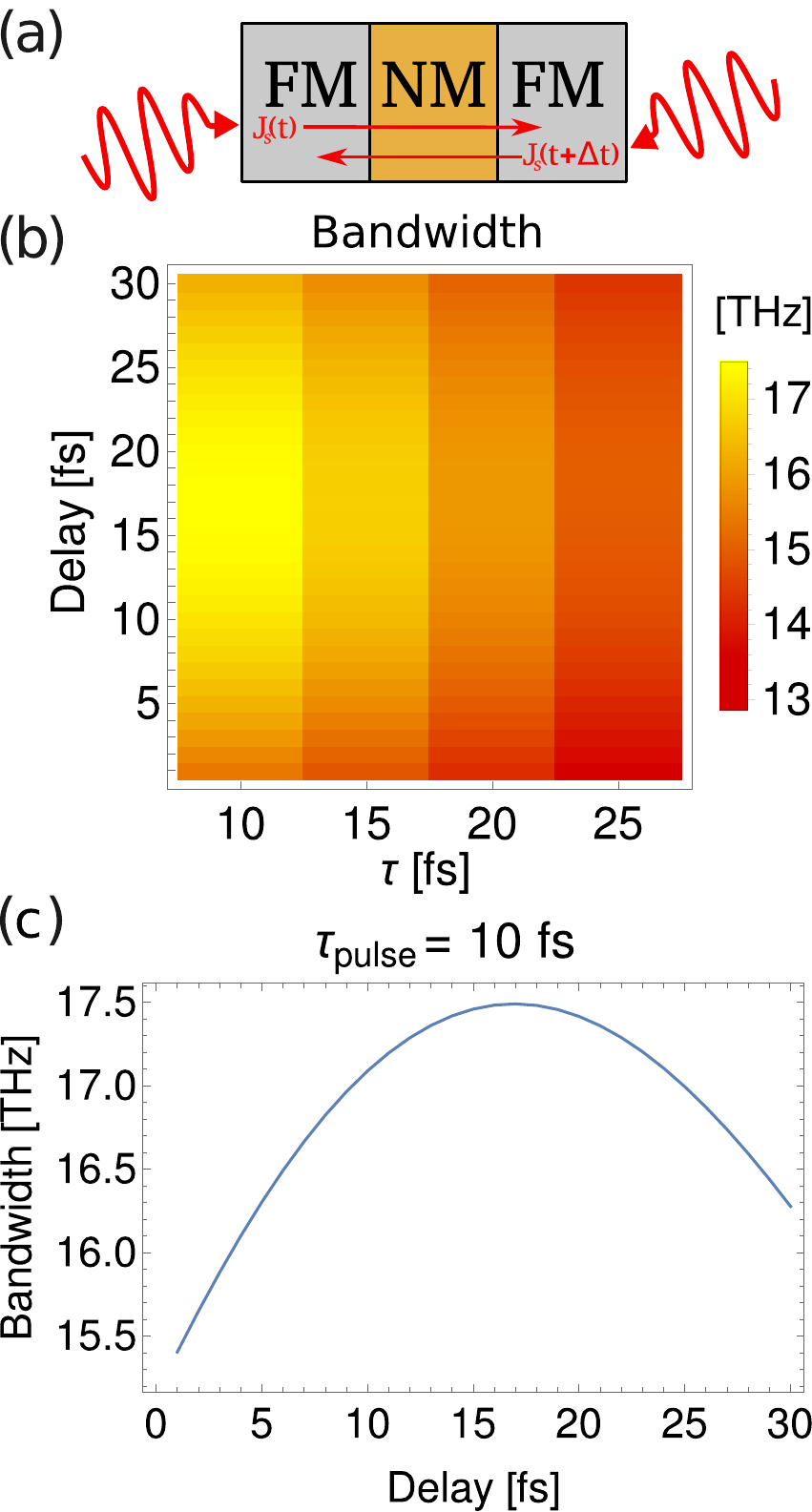}
    \caption{(a) Schematic representation of the two pump-pulses setup. The FM/NM/FM trilayer is excited from both sides by two {time-delayed} laser pulses. The resulting spin currents move in opposite directions and can interfere with each other while propagating in the system. 
    (b) Computed THz bandwidth as a function of the laser pulses' FWHM $\tau$ and time delay ${\Delta t}$ between the two exciting pulses, for an Co(2)/Pt(8)/Co(2) spintronic trilayer emitter. (c) Computed THz signal bandwidth as function of the time delay $\Delta t$ for $\tau=10$~fs.}
    \label{fig:2pulses}
\end{figure}

Lastly, 
to achieve even larger bandwidths, we consider an alternative setup for a spintronic emitter, a {trilayer excited by two pump-laser pulses.} Figure \ref{fig:2pulses}(a) shows {such} trilayer spintronic emitter, where the two outer layers are composed of a FM material, and the central layer is the heavy NM material that generates the charge current and is responsible for the THz emission. This configuration is 
{distinct from} commonly-used trilayer emitters, wherein the central FM layer is sandwiched between two NM layers of opposite spin Hall angle, typically W and Pt \cite{Seifert2016,Kolejak2024,RouzegarPRA2024}. 
In our different configuration, two separate laser pulses excite the emitter from both sides, resulting in two spin currents that originate in either of the FM layers and propagate through the whole trilayer. 

In particular, the choice of thin FM layers favors a well-defined propagation direction of the excited spin currents. When a laser pulse excites a {metal}, the {hot} electrons can travel in all directions with the same probability; however, if we consider a thin layer, electrons that travel toward the outer boundary 
will immediately scatter at the outside interface and be reflected, reverting their direction (we are considering the case of perfectly reflecting boundary conditions, but a spin sink would just {adsorb electrons traveling in one direction} 
and produce similar results), thus resulting in an excited spin current that {flows} mostly in the direction 
of the interface 
with the NM material.

{By exciting with} two laser pulses with a certain delay ${\Delta t}$, the two opposite-traveling {spin-polarized} currents will interfere destructively, resulting in a signal {being} narrower in time and, as such, {wider} in frequency space.  
{Figure \ref{fig:2pulses}(b) shows the THz bandwidth results computed for a Co(2)/Pt(8)/Co(2) trilayer.}
As this figure reveals, the two-pulse excitation can result in a larger or shorter bandwidth, {as compared to the bilayer spintronic emitter}, depending on the delay $\Delta {t}$ between the two pulses {and the pump pulse duration $\tau$}. If the delay between the laser pulses is too short or too long, then the resulting spin currents do not interfere in an optimal way. {The time-delayed second spin current $J_s (t+\Delta t)$ should cancel the tail of the first spin current $J_s(t)$ (cf.\ \cite{Foggetti2025}) to produce a resulting signal more confined in time, according to $J_s(t) - J_s(t +\Delta t) \approx - \Delta t \,dJ_s(t)/dt$.} For a delay of the order of the FWHM of the {initial} pulse, we obtain the maximum value of the bandwidth, see Fig.\ \ref{fig:2pulses}(c). 

As shown in Sec.~\ref{sec:interface}, the properties of the THz emission are strongly dependent on the geometry of the emitter. Hence, a comparison between a trilayer and a bilayer spintronic emitter must be discussed carefully.
In the bilayer case we found an optimal geometry in the Co(2)/Pt(5-6) {bilayer to achieve} the largest emission, {but} this is not the case for a trilayer.
Comparing different configurations of various thickness $x$ of the Pt layer, i.e., Co(2)/Pt(x)/Co(2) (with $x$  ranging from 4 to 14~nm)
{we} found {that} the largest bandwidth {corresponds} to {somewhat} thicker platinum layers, {of about 8 nm}, contrary to the bilayer case.  

In Fig.~\ref{fig:2pulses}(c) we report the results for a Co(2)/Pt(8)/ Co(2) trilayer, and not for the thickest Pt layer we considered. For Pt thicknesses {larger} than 8~nm, the bandwidth {remains practically constant.} 
Hence, we consider the 8~nm layer more interesting as it is the thinnest layer that results in a larger bandwidth than the one of the best bilayer, Co(2)/Pt(6), as reported in Fig.~\ref{fig:size}.
It should additionally be noted that there is a trade-off between the bandwidth and the amplitude of the emitted radiation. 
When studying the total integrated power in the two-pulse setup, we observe a trend very similar to Figure~\ref{fig:2pulses}(b); however, the maximum integrated power in this case is approximately one order of magnitude smaller than the maximum in the bilayer case. This makes the trilayer Co-based system closer in energy efficiency to a Fe/Pt bilayer (see Appendix \ref{app:bilayers}). In fact, depending on the time delay between the two pulses, the temporal squeezing of the spin current can reduce its amplitude, which in turn leads to a smaller peak amplitude.

\section{Discussion and conclusions}

{Our investigation contributes to the on-going discussion on the physical origin of spintronic THz emission. Several mechanisms and models have been proposed to explain such THz emission. Apart from the superdiffusive spin transport \cite{Battiato2012,Kampfrath2013,Balaz2023}, THz emission has been attributed, among others, to the spin Seebeck effect \cite{Seifert2018,Choi2015,Alekhin2017}, spin voltage \cite{Rouzegar2022}, spin pumping \cite{Lichtenberg2022,Beens2023}, and spin currents with diffusive \cite{Dang2020AppPhys} and both ballistic and diffusive transport characteristics \cite{Jechumtal2024}. Recently, also {time-dependent density functional theory} (TD-DFT) has been employed to calculate the charge dynamics and resulting THz electric field \cite{Varela2024,Kefayati2024}.

Several differences and similarities between these models can be observed. The electron transport in the superdiffusive regime and in TD-DFT calculations is nonthermal, whereas other approaches assume thermalized spin-polarized transport. The direction of the spin $\bm{\sigma}$ in the superdiffusive spin current is longitudinal to the moment $\mathbf{M}$ in the FM layer, i.e., $\bm{\sigma} \, || \,\mathbf{M}$. Also in TD-DFT calculations the reduction of the magnetic moment in the FM layer is longitudinal \cite{Varela2024,Kefayati2024}. The spin direction in the conventional spin pumping mechanism \cite{Tserkovnyak2002} is conversely transverse, i.e., $\bm{\sigma} \, || \, (\mathbf{M} \times \partial \mathbf{M}/\partial t)$, {thus, $\bm{\sigma} \perp \bm{M}$}. An additional conversion step due to spin accumulation at the FM/NM interface has been assumed in Refs.\ \cite{Lichtenberg2022,Beens2023} that makes the outgoing spin current longitudinal. This longitudinal nature of the spin current is consistent with direction-sensitive magneto-optical measurements \cite{Rudolf2012,Wieczorek2015,Hofherr2017,Gupta2023}. Further, superdiffusive spin transport contains simultaneously electrons that have not scattered much {during the first tens of fs} and are moving nearly ballistically, as well as electrons that have undergone many scatterings and propagate in the diffusive regime \cite{Battiato2012}. Due to the electron scatterings the amount of ballistic electrons decays with time after the laser excitation. A combination of ballistic and diffusive spin transport \cite{Jechumtal2024} can provide a reasonable approximation to this process at a certain time.}  
{An asset of the superdiffusive transport model is furthermore that it is based on \textit{ab initio} calculated materials' specific quantities. Ultimately, however, the comparison with experiment should resolve what the most applicable model to describe  spintronic THz emission is.} 

{In this work,} we have systematically explored the spectral features of the THz spectrum when ultrafast spin currents, as computed by the superdiffusive spin-transport model, are taken as source terms for the emitted {electric} field.
In our modeling, we take into account the energy dependence of the spin Hall effect in Pt, thus effectively accounting for the {stronger contribution of hot electrons with energies below 0.5 eV, whereas higher-energy electrons with a low value of the spin Hall conductivity do not contribute directly, but only after scattering and losing energy to populate energy levels with higher spin Hall conductivity.} 

Considering a Co/Pt THz emitter as a {typical} test case, {we obtain} very
good agreement  between {numerical} simulations and experimental data for a Co(2)/Pt(4) {bilayer, see Fig.\ \ref{OPT-fig:emissionZT}}. {Both} the THz emission peak and the THz bandwidth are correctly predicted, {which supports that the here-proposed modeling provides a {suitable} description of the ultrafast spin current processes leading to the spintronic THz emission.}

{Having achieved good agreement between our simulations and experiment, we have aimed to identify optimal conditions to maximize the THz emission. To this end, we have}
explored {numerically} the effect of different pulse durations, the {influence of} the relative thicknesses of the two composing materials, the role of the reflecting or transmitting properties of the FM/NM interface, as well as  of the cap and substrate layers. {In addition, we have proposed and simulated a {trilayer} emitter in a double pulse setup}. 

{To summarize,} our simulations show specifically, first, that shorter excitation pulses are favorable to reach larger bandwidths and higher values of the peak emission frequency. 
Second, 
our simulations show that a thin heavy-metal layer ($5-6$~nm)
provides a sizable bandwidth and good peak emission frequency as compared to thicker nonmagnetic NM layers.
Similarly, thin FM layers ($2 -3$ nm) are predicted to provide good THz pulse bandwidths and peak frequencies. 
Third, {investigating the transmission properties of} the FM/NM interface {we find that, for obtaining a large bandwidth, the interface should optimally} act as a filter {with a high transmissivity of} a hot electrons injected into the NM layer, but {with a low transmission of hot electrons flowing back into the FM layer, so}
that electrons are confined longer in the NM side of the emitter. 
Fourth, modeling the transmission/reflection at external boundaries of bilayer THz emitters, we find that larger bandwidths and higher emission frequencies are expected for reduced reflections at the boundaries, {i.e., for spin sinks,} however, with a trade off for the {emitted power} or the maximum amplitude of the generated signal {being reduced.}  Finally, {we have proposed} an alternative way to increase the bandwidth of the emitted signal {by using} a double-pulse setup that could {become} a versatile tool in actively tuning the emitted {electric} field while leaving the emitter itself unchanged. {The here-predicted THz emitter properties can be examined in dedicated experiments, which will further contribute to establishing rigorously the {disputed} mechanisms that lead to THz electric field emission.}

\begin{acknowledgments}
 This work has been supported by the Swedish Research Council (VR), the German Research Foundation (Deutsche Forschungsgemeinschaft) through CRC/TRR 227 ``Ultrafast Spin Dynamics'' (project MF, project-ID: 328545488)), the K.\ and A.\ Wallenberg Foundation (Grants No.\ 2022.0079 and 2023.0336), {the French National Research Agency PEPR SPIN ANR-22 EXSP-0003 ``TOAST'',} {and French National Research Agency (ANR) under ANR-21-CE24-0011 ``TRAPIST''.}
 {This work was further supported by} the European Union’s Horizon 2020 Research and Innovation Programme under FET-OPEN Grant Agreements No.\ 863155 (s-Nebula) {and  964735  (EXTREME-IR)}. The computational resources were provided by the National Academic Infrastructure for Supercomputing in Sweden (NAISS) at NSC 
Link\"oping, partially funded by VR through Grant Agreement No.\ 2022-06725.
\end{acknowledgments}

\section*{Data Availability Statement}
The data that support the findings of this article are publicly available \cite{data_zenodo}.

\appendix

\begin{figure*}[!bth]
\begin {center}
\includegraphics[width=.9\linewidth]{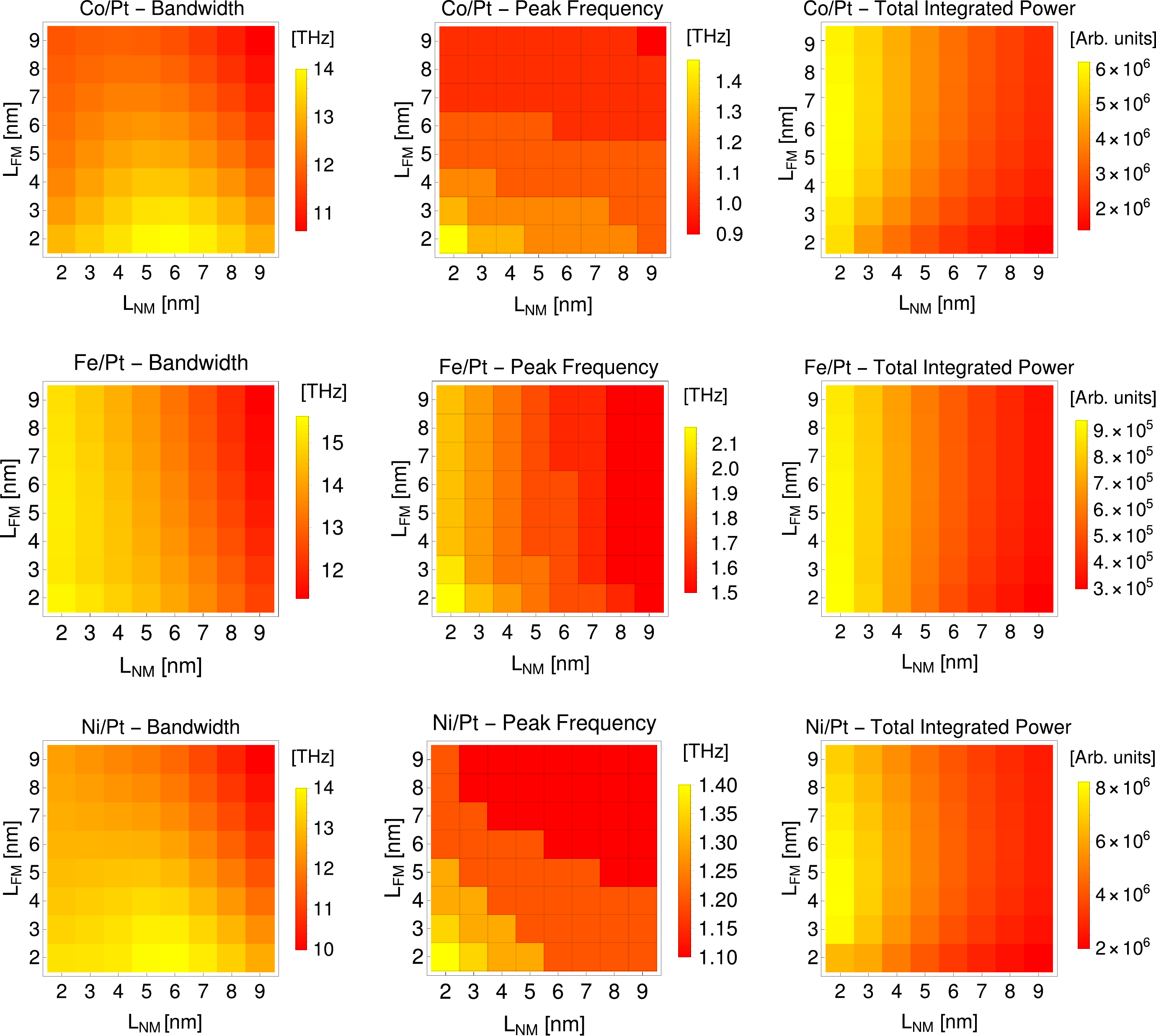}
\end{center}
\vspace*{-0.4cm}
\caption {
{Calculated THz}  signal bandwidth (left), THz peak frequency (center), and total integrated power of the emitted radiation (right), plotted as a function of the FM and NM layer {thicknesses},  L$_\mathrm{FM}$ and L$_\mathrm{NM}$. {The THz emission properties are calculated for} a Co/Pt (top row), Fe/Pt (middle row), and Ni/Pt (bottom row) bilayer, respectively.
\label{sfig:thicknessFeCoNi}
}
\end{figure*}

\section{Influence of layer thicknesses for Co/Pt, Fe/Pt and Ni/Pt}
\label{app:bilayers}

Here we report additional {results} for the {simulated THz emission of} Co/Pt, Fe/Pt and Ni/Pt bilayers.
{In Fig.\ \ref{sfig:thicknessFeCoNi} we show the calculated bandwidth, peak frequency, and total integrated power, respectively, as a function of the layer thicknesses $\mathrm{L_{NM}}$ and $\mathrm{L_{FM}}$.} As in Fig.~\ref{fig:size}, we set the boundary reflection coefficients at 0.95 when computing the peak frequency. 
Qualitatively, changing the FM metal of the bilayer does not change dramatically the THz properties of the emitter. We {observe for each of the 3$d$ FMs} a preference for {thin} FM layers {to achieve a large bandwidth,} {together} with an intermediate  thickness of the Pt layer.\\

Similar to the Co case, a maximum bandwidth is, {for example}, obtained for the Ni(2)/Pt(6) {bilayer}, while Fe seems to perform better with the thinnest Pt layer. The dependence of the peak frequency {on} the emitter {thickness} is weak {for each of the 3$d$ FMs}, and {also} the total integrated power is qualitatively the same. It is worth noting that a higher integrated power is obtained for thicker FM layers, {because these lead to} a greater demagnetization current and a higher injection of spin-polarized electrons into the NM layer.  

\begin{figure*}[!tbh]
    \centering
    \includegraphics[width=0.94\textwidth]{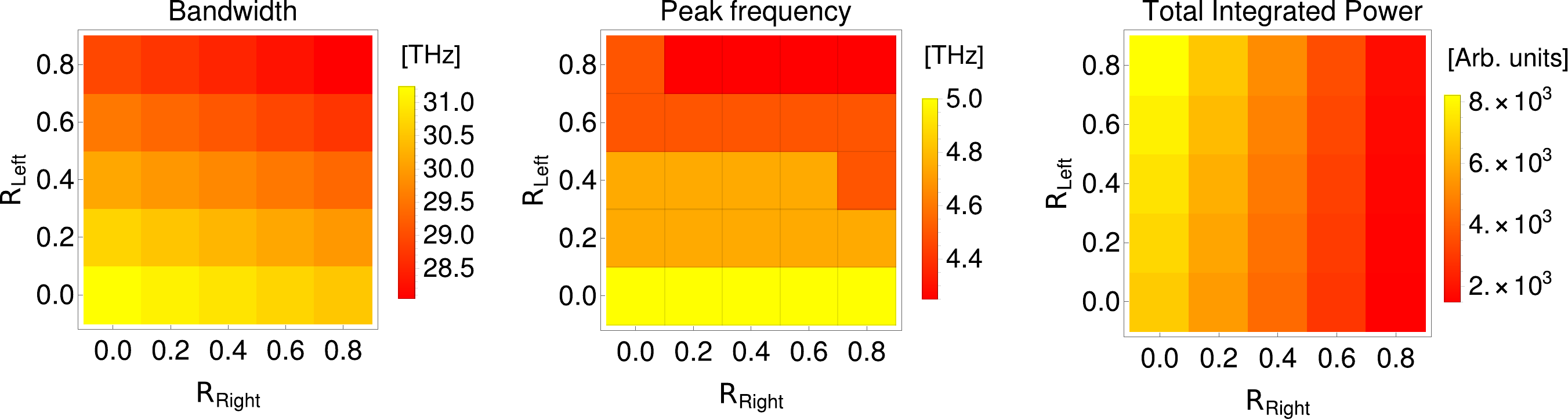}
    \vspace*{-0.3cm}
    \caption{
    Simulated THz signal bandwidth, peak frequency, and total integrated THz power as a function of the interface reflection coefficients for left- and right-moving electrons, $\mathrm {R_{Left}}$ and $\mathrm {R_{Right}}$, respectively. The signals are {calculated for an} Co(2)/Pt(4) bilayer with perfectly transparent outer boundaries ($\mathrm{R=0}$), {i.e., the outer layers act as a hot electron spin sink.}}
    \label{sfig:filter0}
\end{figure*}

\section{Interface effects combined with nonreflective boundaries}\label{App:interfaces}

{In Sec.\ \ref{sec:interface} we considered the influence of reflection and transmission at the FM/NM interface, in combination with fully reflective outer boundaries, i.e., R$_{\mathrm{FM}} = \mathrm{R_{NM}}=1$, typical for a spintronic emitter on an insulator substrate.}
Here we report additional {simulations for an Co(2)/Pt(4) emitter with nonreflective outer boundaries.}  
Figure~\ref{sfig:filter0}  {shows the computed THz signals}  when we consider the outer boundaries to be completely transparent, i.e., $\mathrm{R} =0 $, a hot electron and spin sink. Thus, the electrons that reach the outer boundaries  escape the system. {Under this condition} we {observe} that the most important {quantity} is the reflection coefficient R$_\mathrm{Right}$ which acts as a filter for the electrons being injected into the NM layer. The lower the value of R$_\mathrm{Right}$, the higher {is} the number of {injected} electrons contributing to the emission and, as a consequence, the higher {is} the bandwidth, peak frequency, and  total integrated THz power.

The results are quite different from Fig.~\ref{fig:filter}. In Fig.~\ref{sfig:filter0} the dependence of the bandwidth and peak frequency on R$_\mathrm{Right}$, which is related to the probability of right-moving electrons to be reflected at the Co/Pt interface, is rather weak and these THz quantities are much more sensitive to changes of R$_\mathrm{Left}$. conversely, the total integrated power is found to be mostly independent of R$_\mathrm{Left}$ and more sensitive to changes in R$_\mathrm{Right}$. 

We mention however that the total change in bandwidth, when varying the reflection coefficients, of approximately few THz, is relatively small when compared with a bandwidth of the order of 30 THz. 
The { explanation of}  this behavior is that, for transparent outer boundaries, there {are no electrons reflected at the outer boundary of the} NM(FM) layer and thus the left-(right-)moving current is almost non-existent, with the exception of a {limited} number of electrons that are randomly reflected back from inside the thin Pt(Co) layer. Hence, the effect of R$_\mathrm{Left}$(R$_\mathrm{Right}$) is negligible when the outer boundaries act as  a perfect spin sink (R$_{\rm Boundary}=0$).

\begin{figure}[t!]
    \centering
    \includegraphics[width=.36\textwidth]{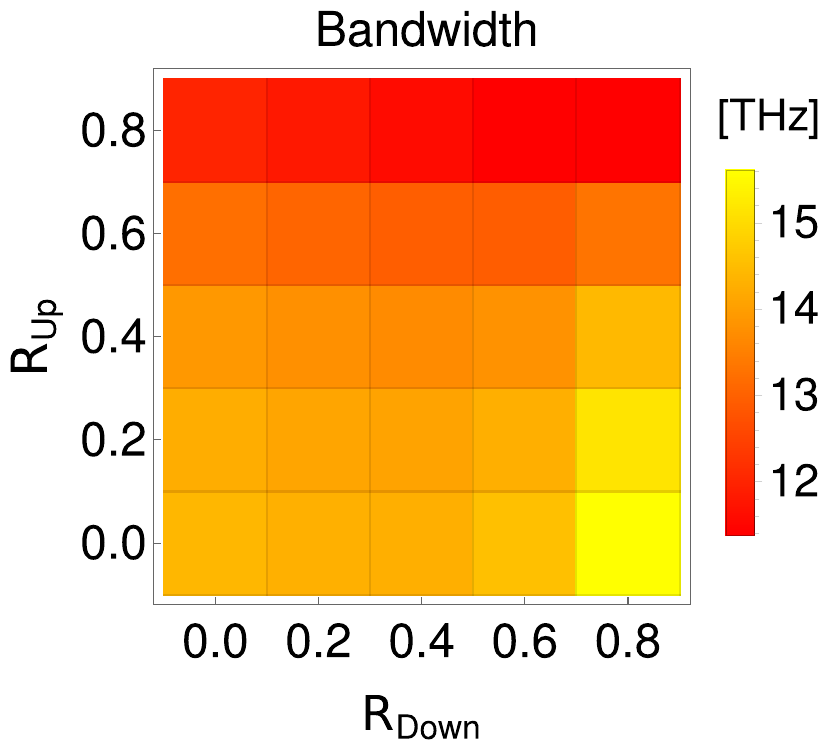}
    \vspace*{-0.3cm}
    \caption{{Simulated} bandwidth of an Co(2)/Pt(4) spintronic emitter as a function of the Co/Pt interface reflection coefficients for spin up ($\mathrm{R_{Up}}$) and spin down ($\mathrm{R_{Down}}$) electrons. The reflection coefficients are assumed to be independent of the electrons' energy.}
    \label{sfig:UpDown}
\end{figure}

Next, we consider the influence of different reflection/transmission coefficients for spin majority and minority electrons. Recent \textit{ab initio} calculations have shown that such difference can be expected to occur \cite{Dang2020AppPhys,Balaz2023}. We consider the Co(2)/Pt(4) bilayer and assume that the reflection coefficients for left and right-moving electrons are the same, i.e., $\mathrm{R= R_{Left} = R_{Right}}$ and we do not include an energy dependence of the reflection coefficient, but we assume different reflection coefficients for spin up ($\mathrm{R_{Up}}$) and spin down
($\mathrm{R_{Down}}$). The calculated result for the effect of spin-dependent reflection coefficients on the THz bandwidth is shown in
Fig.\ \ref{sfig:UpDown}. 
{One can straightforwardly} observe that the maximum of the bandwidth is obtained for a low reflection of the spin-up electrons, while the reflective properties of the interface for the spin-down electrons has {only a minor influence} on the bandwidth. This {highlights} how the majority spin carriers {injected from the FM} give the biggest contribution to the THz emission. 

We note additionally that the results of the spin sink and interface dependent configurations (cf.\ Figs.~\ref{fig:filter}, \ref{sfig:filter0} and \ref{sfig:UpDown}) do not depend strongly on the choice of FM metal. We performed the calculations with Fe and Ni instead of Co and, aside from different bandwidth values, the qualitative behavior is the same.

{
Lastly, we return to the energy dependence of the reflection/transmission at the FM/NM interface. The reflection of hot electrons at the interface can depend on the energy of the hot electrons, a situation which, according to \textit{ab initio} calculations, frequently occurs \cite{Dang2020AppPhys,Balaz2023,Lu:PRB2020}. In Fig.\ \ref{fig:filter} we showed  the computed bandwidth for different values of the interface reflection coefficients {in an Co(2)/Pt(4) bilayer}. In the following we investigate further the influence of the energy-dependent reflectivity {coefficient.}}

{To this end, we change} the reflection coefficients in an energy-dependent fashion. As the reflection coefficients must only assume values between 0 and 1, it is usually not possible to increase, or decrease, uniformly the different coefficients at different energies without exceeding the interval [0,\,1].  Hence,  to increase, or decrease, the reflection coefficient in a way that maintains the energy-dependent character of the coefficients we define the following quantities:
\begin{equation}
\begin{split}
&\mathrm{R_{decr}}(E)=\mathrm{R}(E)-\alpha \mathrm{R}(E) ,\\
&\mathrm{R_{incr}}(E)=\mathrm{R}(E)+\beta (1-\mathrm{R}(E)),
\end{split}
\label{seq:Rid}
\end{equation}
where $\alpha$ and $\beta$ {can be seen} to {decrease} and {increase}, respectively, the reflectivity, and are both positive and smaller than 1. It follows that $\mathrm{R_{decr}}$ is smaller or equal to $\mathrm{R}(E)$, but never negative, as well as $\mathrm{R_{incr}}$ is always greater or equal to $\mathrm{R}(E)$, but never exceeds 1. By using $\mathrm{R_{decr}}(E)$ or $\mathrm{R_{incr}}(E)$ instead of the usual $\mathrm{R}(E)$ (taken from Ref.\ \cite{Lu:PRB2020}),  
we can study how the bandwidth changes with decreasing (increasing) $\alpha$ or increasing {(decreasing)} $\beta$ values of the interface reflection coefficient. This approach allows to retain the energy-dependent character of the reflection coefficients, {but we must be aware that the increase or decrease of the reflection coefficients depends on the energy and on the original values of the coefficients}. 
%
\begin{figure}[!t]
\vspace*{0.4cm}
    \includegraphics[width=.4\textwidth]{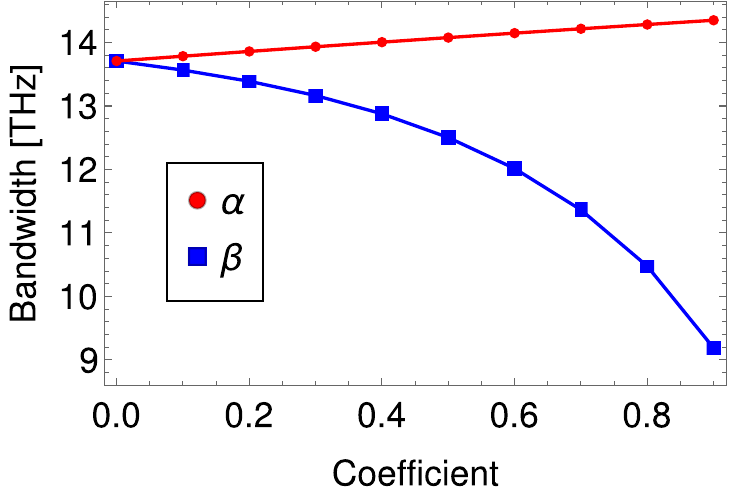}
    \vspace*{-0.2cm}
    \caption{{Calculated} variation of the THz bandwidth {of a Co(2)/Pt(4) emitter} as function of the coefficients $\alpha$ and $\beta$, see
   Eq.\ (\ref{seq:Rid}).  The values of the interface reflection coefficients $\mathrm{R} (E)$ are reduced, for every value at different energies,  
   for larger coefficient $\alpha$, {whereas they are increased} 
   {for larger coefficient} $\beta$. }
    \label{sfig:E-dep}
\end{figure}
Figure~\ref{sfig:E-dep} shows the result of {such simulations for the Co(2)/Pt(4) emitter}. {It can be seen that} by reducing the reflection coefficients at the interface {(larger $\alpha$)} the bandwidth increases, similarly to the uniform case described in the main text. Conversely, as expected, an increase of the reflection coefficient {(larger $\beta$)} generates a reduction of the bandwidth, further corroborating the results reported in Sec.\ 
\ref{sec:interface}  for the simplified case of uniform reflection coefficients. This behavior does not depend on the choice of the FM metal.

\section{Reflection coefficients}
\label{app:ref_coeff}
\begin{figure}[h!]
    \centering
    \includegraphics[width=0.75\linewidth]{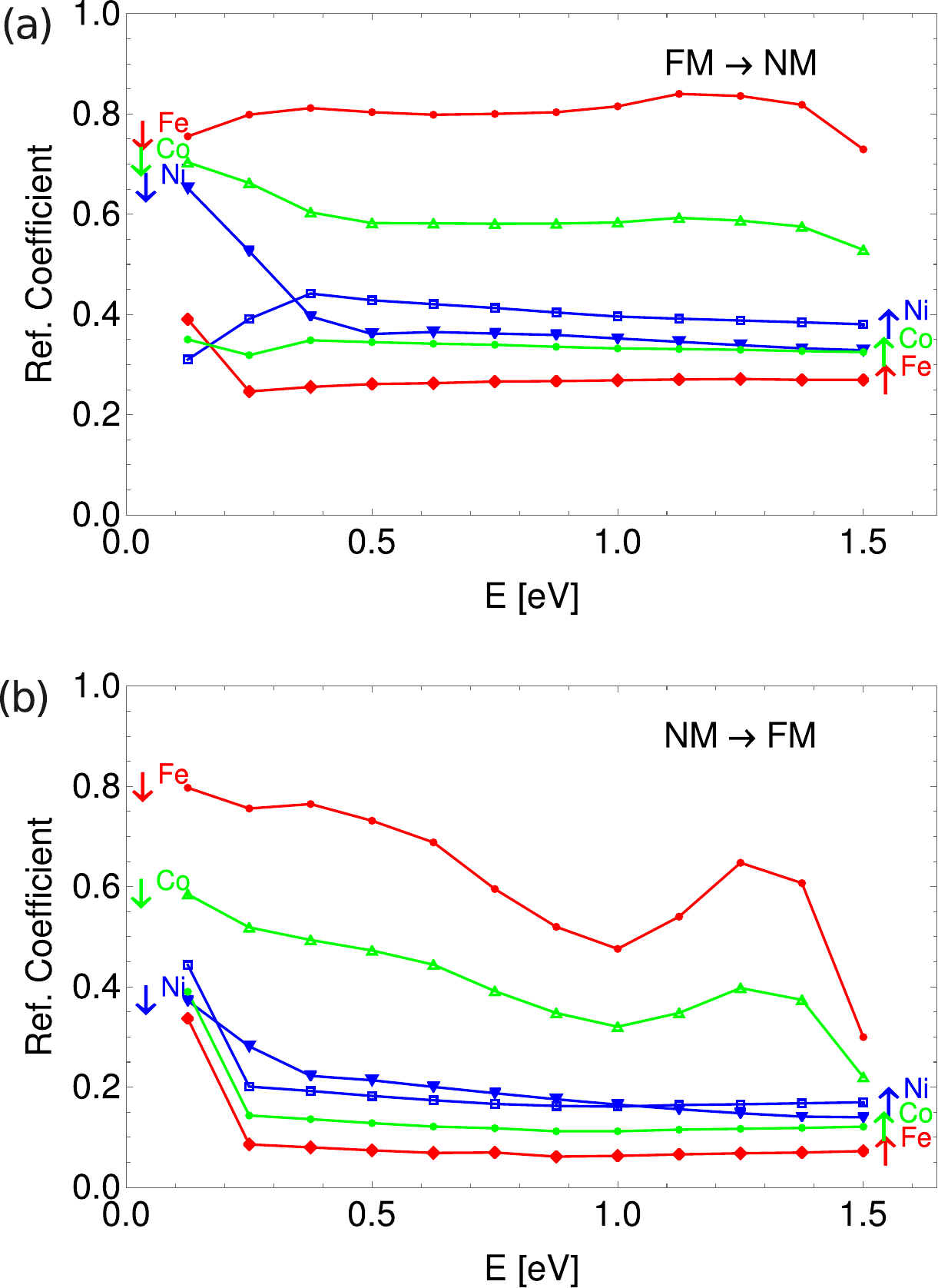}
    \vspace*{-0.3cm}
    \caption{Reflection coefficients taken from Ref.\ \cite{Lu:PRB2020} for Fe/Pt and Ni/Pt interfaces as a function of energy for spin up ($\uparrow$) and down ($\downarrow$) electrons, traveling from the FM to the Pt NM layer (a) and \textit{vice versa} (b). 
    We assume the reflection coefficients for Co/Pt  as an average of the coefficients for 
    Fe/Pt and Ni/Pt.}
    \label{sfig:reflections}
\end{figure}

In Fig.~\ref{sfig:reflections} we report the reflection coefficients we used in our simulations. The coefficients for Ni/Pt and Fe/Pt interfaces are taken from Ref.\ \cite{Lu:PRB2020}. The coefficients for Co/Pt are not available, so we estimated them by averaging the coefficients for Fe/Pt and Ni/Pt. 

\FloatBarrier

\bibliography{bibliography}
\end{document}